\documentclass[sigconf,natbib=true]{acmart}
\AtBeginDocument{%
  }

\copyrightyear{2026}
\acmYear{2026}
\setcopyright{cc}
\setcctype{by}
\acmConference[ICTIR '26]{Proceedings of the 2026 International ACM SIGIR Conference on Innovative Concepts and Theories in Information Retrieval (ICTIR)}{July 25, 2026}{Melbourne, VIC, Australia}
\acmBooktitle{Proceedings of the 2026 International ACM SIGIR Conference on Innovative Concepts and Theories in Information Retrieval (ICTIR) (ICTIR '26), July 25, 2026, Melbourne, VIC, Australia}
\acmDOI{10.1145/3805713.3820429}
\acmISBN{979-8-4007-2600-2/2026/07}

\usepackage{booktabs}

\usepackage{amssymb}
\usepackage{booktabs}
\usepackage{graphicx}
\usepackage{tabularx}
\usepackage{titlesec}
\usepackage{amsmath}
\usepackage{booktabs}
\usepackage{multirow}
\usepackage{xurl}  % better URL line breaking
\usepackage{soul} 
\usepackage{lineno}
\usepackage[skip=2pt]{caption}
\usepackage{wrapfig}
\usepackage{float}
\usepackage{lineno}
\usepackage{bm}
\usepackage{float} % in preamble for [H]
\usepackage{subcaption} % if you want subfigure captions
\usepackage{placeins}
\usepackage{booktabs} % for cleaner tables (optional but recommended)
\definecolor{darkblue}{rgb}{0, 0, 0.5}
\hypersetup{colorlinks=true, citecolor=darkblue, linkcolor=darkblue, urlcolor=darkblue}

\titlespacing*{\section}{0pt}{*1}{*1} % Adjusts spacing before and after \section
\titlespacing*{\subsection}{0pt}{*1}{*1} % Adjusts spacing for \subsection

\begin{document}

\title{LLM-Driven Usefulness Judgment for Web Search Evaluation}

\author{Mouly Dewan}
\affiliation{%
  \institution{Information School\\ University of Washington}
  \city{Seattle, WA}
  \country{United States}}
\email{mdewan@uw.edu}

\author{Jiqun Liu}
\affiliation{%
  \institution{The University of Oklahoma}
  \city{Norman, OK}
  \country{United States}}
\email{jiqunliu@ou.edu}

\author{Aditya Gautam}
\affiliation{%
  \institution{Meta}
  \city{Bellevue, WA}
  \country{United States}}
\email{agautam7041@gmail.com}

\author{Chirag Shah}
\affiliation{%
  \institution{Information School\\ University of Washington}
  \city{Seattle, WA}
  \country{United States}}
\email{chirags@uw.edu}
\begin{abstract}
Evaluation is fundamental to optimizing search experiences and supporting diverse user intents in Information Retrieval (IR). Traditional search evaluation methods primarily rely on relevance labels, which assess how well retrieved documents match a user's query. However, relevance alone fails to capture a search system’s effectiveness in helping users achieve their goals, making usefulness a critical evaluation criterion. Recent LLM-enabled evaluation works have mostly focused on relevance label generation. In this paper, we explore an alternative approach: LLM-generated usefulness labels that incorporate implicit and explicit user behavior signals along with relevance. We introduce Task-aware Rubric-based Usefulness Evaluation (TRUE), a reproducible rubric-driven framework that leverages iterative sampling and Chain-of-Thought reasoning to model complex search behavior patterns. Our study shows that: (i) pre-trained LLMs can generate moderate usefulness labels with rich session-level context; and (ii) LLMs with TRUE outperform state-of-the-art methods and our systematically constructed baseline. We further analyze the relationship between usefulness and user satisfaction by comparing label generation with and without satisfaction signals, quantifying their correlation and impact on model behavior. Additionally, we conduct an ablation study to identify key features for accurate usefulness label generation, enabling cost-effective evaluation. Overall, this work advances LLM-based evaluation beyond relevance by proposing a reproducible and scalable framework for usefulness judgment, addressing key reproducibility challenges.
\end{abstract}




\maketitle

\section{Introduction}
In IR evaluation scenarios, assessing relevance alone is not enough, as it often fails to capture the extent to which a document is \textit{useful} for an underlying task~\cite{cole2009usefulness, belkin2008relevance}. While prior work argues that relevance is insufficient for evaluating search systems~\citep{cole2009usefulness, liu2022toward}, mainstream IR has largely focused on objective relevance assessment. Traditional Cranfield paradigms \citep{cormack1998efficient} evaluate documents individually, ignoring broader task goals \citep{belkin2008relevance} and limiting applicability to interactive information retrieval. Relevance annotations alone cannot fully represent document usefulness or user satisfaction \citep{10.1145/2766462.2767854, belkin2009model, mao_when_2016}, and third-party annotators often lack users’ situational context, creating a \textit{situational disconnect}. Prior studies show weak correlations between relevance metrics and user satisfaction \citep{mao2016does}. \textit{Usefulness judgment} is deeply context-dependent, carrying richer contextual information than relevance but is also more challenging to capture in real time or reuse in evaluation experiments~\citep{liu2022leveraging, jiang2026improving}. Therefore, our work incorporates usefulness as an evaluation metric to provide a more holistic assessment, capturing not only document relevance but also the user’s search intent and satisfaction throughout the search session~\citep{mitsui2017predicting, liu2020identifying, liu2021state}.

% \begin{figure}[t]
% \centering
%     \includegraphics[width=\linewidth]{true_reasoning.png}
%     \captionsetup{aboveskip=15pt, belowskip=0pt}
%     \caption{Usefulness label generation with LLMs.}
%     \label{fig:usefulness_generation}
% \end{figure}

Moreover, traditional IR evaluation has relied on human assessors and crowd workers, which is time-consuming and costly \citep{thomas_large_2024}. Human evaluations are also difficult to reproduce and often not directly comparable across rounds due to inconsistent annotator pools~\citep{jiang2026improving}. To address these challenges, researchers have explored automated evaluation using large language models (LLMs)~\citep{dietz2025principles}. LLM-based evaluation (LLM4eval) \citep{rahmani_llm4eval_2024} has gained traction due to its strong reasoning and task-solving capabilities \citep{wei2022chain, Wei2022EmergentAO}. However, most LLM-enabled IR evaluation efforts focus on TREC-style document relevance~\citep{rahmani_llmjudge_2024, chen2026mitigating}, while user-centric usefulness remains underexplored. Therefore, this paper evaluates LLM-generated usefulness labels that incorporate search session, task, query, and document-level user signals to better capture user perceptions of system quality. However, usefulness judgment poses challenges beyond those encountered in relevance evaluation. First, it is unclear how to effectively represent users' search journeys to LLMs. Second, the choice of features and criteria that should guide usefulness prediction remains under-specified. Finally, LLM-based evaluation itself raises reproducibility concerns, as model outputs can vary across prompt formulations, inference settings, and model versions. Addressing these challenges requires the design of an evaluation framework specifically tailored to usefulness prediction.

We address these challenges by incorporating both implicit and explicit user behavior signals into LLM-based usefulness evaluation. Implicit signals include Clicks, Click-Through Rate (CTR), Query Dwell Time, Task Dwell Time, and URL Dwell Time capture engagement, while explicit signals include task satisfaction, query satisfaction, task relevance, and query relevance. We also incorporate document-level features such as URL, title, summary from the search engine results page (SERP), ranking position, associated queries, and query position. These signals enable richer representations of users’ search journeys and provide concrete criteria for usefulness assessment. While satisfaction may not be available in real-world deployment and can be considered downstream of usefulness, we treat it as a post-hoc signal in an offline evaluation setting and explicitly analyze its impact on usefulness estimation. Most existing LLM-based judgment methods rely on prompt-based procedures that are highly sensitive to prompt design, often limiting reproducibility \cite{10.1145/3731120.3744591}. To address this, we introduce TRUE (Task-aware Rubric-based Usefulness Evaluation), a rubric-driven framework that structures LLM reasoning over search context and behavioral signals. We compare TRUE with state-of-the-art LLM methods, traditional machine learning models, our own curated baseline, and session-aware methods using \textit{Chain-of-Thought} (CoT) \citep{wei2022chain} reasoning. This work is guided by the following RQs:
\begin{itemize}
    \item \textbf{RQ1:} Can LLMs generate reliable usefulness judgments when provided with representations of search sessions?
    \item \textbf{RQ2:} Does a rubric-driven evaluation framework improve the performance of LLM-based usefulness assessment?
    \item \textbf{RQ3:} What is the effect of user satisfaction signals on LLM-based usefulness judgments?
    \item \textbf{RQ4:} What are the most effective features in improving LLMs’ document usefulness judgments?
\end{itemize}

Motivated by these research questions, this work examines whether LLMs can assess usefulness beyond traditional relevance-based evaluation addressing key limitations in current IR methods. The \textbf{novelty} lies in introducing \textit{TRUE (Task-aware Rubric-based Usefulness Evaluation)}, a rubric-driven reproducible method that integrates implicit and explicit user signals to evaluate document usefulness. The main \textbf{contributions} of our work include:

\begin{itemize}
\item We propose \textbf{TRUE}, a structured and reproducible rubric-based framework for usefulness evaluation using LLMs.

\item We introduce a method to represent \textbf{user search sessions} by structuring implicit and explicit behavioral signals into LLM-compatible inputs.

\item We identify the most effective implicit and explicit user signals through an \textbf{ablation study}.

\item We analyze the impact of \textbf{user satisfaction} on usefulness judgments, evaluating robustness with and without them.
\end{itemize}

\section{Background and Related Works}
Evaluation of search systems has long been central to IR research, with the goal of supporting and understanding users' information needs. Traditional evaluation campaigns such as TREC, NTCIR, CLEF, and FIRE \citep{Carmel2010EstimatingTQ, 10.1145/1842890.1842906, jaech-ostendorf-2018-personalized, maddalena2016crowdsourcing} follow the Cranfield paradigm \citep{cormack1998efficient}, focusing on document-level relevance judgments. However, these methods primarily assess individual results and often overlook the broader context of search tasks \citep{belkin2008relevance, liu2021deconstructing}. 

\textbf{Relevance and usefulness judgments} are traditionally obtained through human batch annotation, as in the deep learning tracks of TREC \citep{voorhees2000overview}. Alternatively, relevance can be inferred from user behavior logs \citep{joachims2017accurately} or synthesized using LLM-based methods \citep{bonifacio2022inpars, dai2022promptagator}. Online user metrics such as CTR, UCTR \citep{chuklin2013click}, PLC \citep{chapelle2009expected}, and dwell time \citep{jiang2015understanding} further capture implicit feedback and provide valuable signals for understanding user engagement. While relevance has been the dominant evaluation metric, many researchers argue that it falls short of capturing true system effectiveness in interactive IR. \cite{cole2009usefulness} and others \citep{mao2016does, yilmaz_relevance_2014, zhang_models_2020} emphasize the importance of usefulness in reflecting how well documents help users achieve their task goals. Studies show that relevance and usefulness may diverge, especially when users exert significant effort to extract value from content \citep{yilmaz_relevance_2014}. Because usefulness is more user-centric, it often requires richer behavioral data and context to measure accurately. Replacing relevance-based measures with usefulness-based measures in IR evaluation has also been advocated by \cite{belkin2009model} and \cite{cole2009usefulness}. Automated usefulness labeling remains challenging, particularly as users reformulate queries and adapt their search strategies. Our exploration of LLM usefulness judgment addresses this challenge by incorporating both implicit behavior signals and task context.

\textbf{LLMs in IR Evaluation} has primarily been used to automate relevance labeling, while user-centric usefulness metrics remain relatively underexplored. \cite{faggioli_perspectives_2023} were among the earliest to investigate the use of LLMs for relevance judgment, providing insights into human--LLM collaboration through direct prompting strategies. Another prominent work by \cite{thomas_large_2024} showed that LLMs can predict searcher preferences and produce relevance labels comparable to those of human annotators. Prior studies also show that LLMs tend to be more positive in assessing relevance and should therefore be used to evaluate all documents rather than only relevance gaps \citep{abbasiantaeb_can_2024}. Some work has also examined whether LLMs can perform utility judgment using retrieval-augmented generation (RAG) \citep{zhang_are_2024, salemi2024evaluating}. To address limitations in \cite{thomas_large_2024}, \cite{upadhyay_large-scale_2024} conducted large-scale relevance assessment on TREC Deep Learning tracks using UMBRELLA \citep{upadhyay_umbrela_2024}, a relevance assessor that uses LLMs for academic TREC-style evaluation. Most LLM-based IR evaluation methods still focus on relevance labeling \citep{dewan2026true}, even though relevance alone is insufficient because usefulness more directly reflects user goal achievement \citep{belkin2008relevance, cole2009usefulness}. In this work, we focus on usefulness as a key metric for evaluating web search alongside relevance. To do so, we generate usefulness labels using LLMs by incorporating implicit user behavior signals such as CTR and dwell time \citep{chuklin2013click, chapelle2009expected, jiang2015understanding}. Recent work has begun exploring LLM-based usefulness evaluation. For example, \citet{dewan2025llm} investigate LLM-based usefulness labeling and examine the distinction between relevance and usefulness, while CLUE \citep{wang2025clue} incorporates search context and behavior for usefulness judgment. These approaches represent early steps toward LLM-based usefulness evaluation, and our study directly compares against CLUE.
\begin{figure*}[!t]
  \centering
  \includegraphics[width=0.85\linewidth]{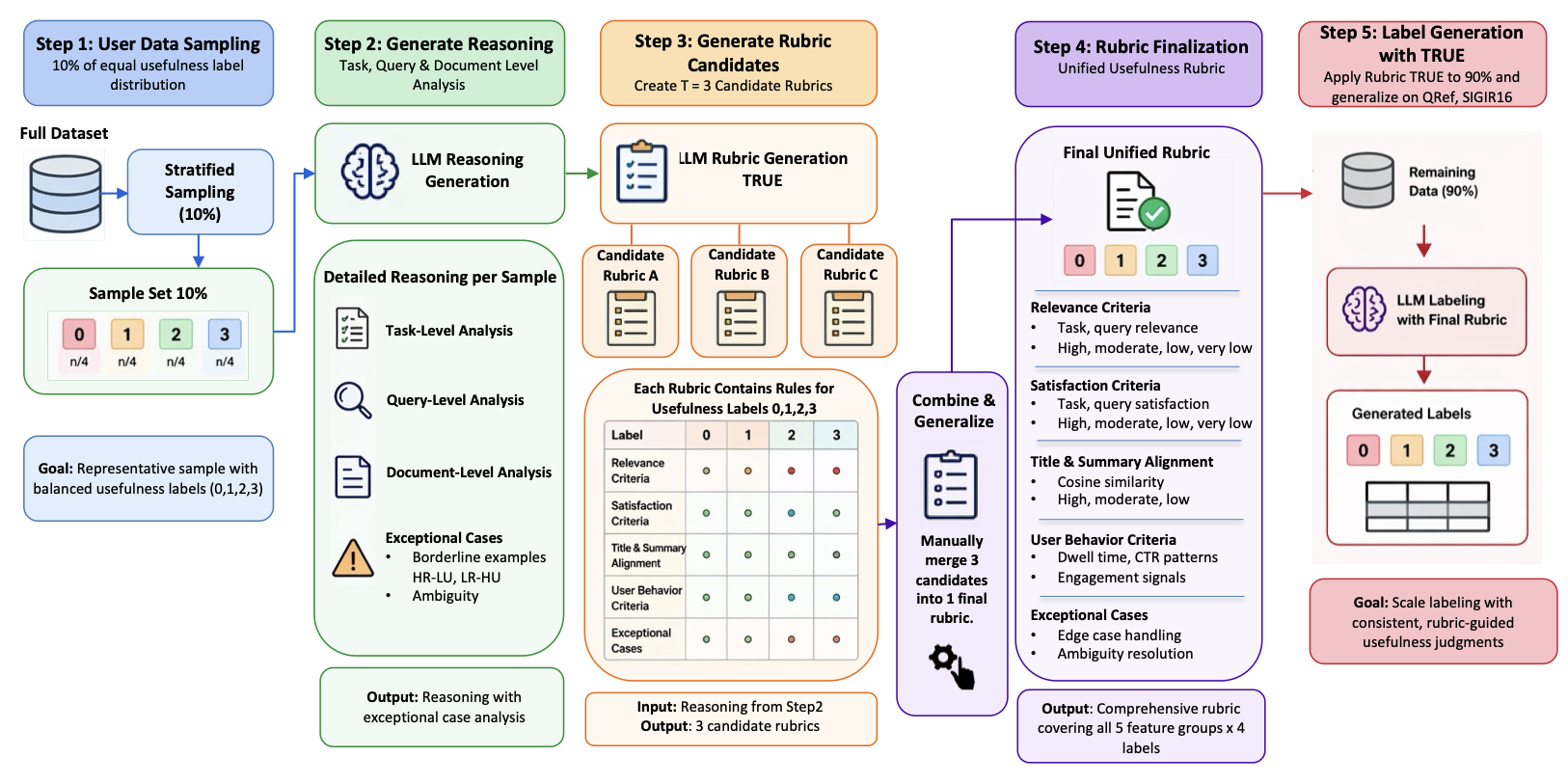}
  \captionsetup{aboveskip=5pt, belowskip=0pt}
  \caption{TRUE formulation (Step 1-5).}
  \label{fig:true_formulation}
\end{figure*}

\vspace{-4pt}
\section{Problem Definition}
We focus on usefulness judgment over clicked documents, as prior work assumes that usefulness is only observable for documents a user interacts with \citep{mao_when_2016}. Unlike relevance, usefulness is highly situational and depends on user behavior and search context \citep{mao_understanding_2017}. Accordingly, our formulation incorporates user behavior signals, contextual information, and explicit measures summarized in Table~\ref{tab:scales_key_measures} in addition to the query–document pair.

Formally, given a query $q$, the set of clicked documents $D_c$, and auxiliary information $i \in I$ capturing implicit and explicit user signals, the task is to predict multilevel usefulness labels $U_{\text{llm}}$:
\begin{equation}
F_U :
\{(q, d, i) \mid d \in D_c,\ i \in I\}
\;\rightarrow\;
\mathcal{U},
\quad \mathcal{U} = \{0,1,2,3\}
\label{eq:usefulness_task}
\end{equation}

Here, $I$ includes implicit behavioral signals (e.g., CTR, query, task, and URL dwell time), explicit feedback (e.g., satisfaction and relevance), and document-level context (e.g., title, summary, rank, and query position). The output of $F_U$ corresponds to $U_{\text{llm}}$, the usefulness labels predicted by LLMs which we evaluate by comparing them against $U_u$, the usefulness judgments provided by real users.

\begin{table}[t]
\small
\centering
\caption{Scales of key evaluation measures across datasets.}
\label{tab:scales_key_measures}
\setlength{\tabcolsep}{3pt}
\renewcommand{\arraystretch}{1.05}
\begin{tabular}{lcccc}
\toprule
\textbf{Variable} & \textbf{KDD'19} & \textbf{QRef} & \textbf{SIGIR16} & \textbf{Annotator} \\
\midrule
Relevance ($R_a$)      & 0-3 & -- & 0-4 & third-party \\
Usefulness ($U_a$)   & 0-3 & 0-3 & 0-3 & third-party \\
Usefulness ($U_u$)*   & 0-3 & 0-3 & 0-3 & real users \\
Usefulness ($U_{\text{llm}}$)+ & 0-3 & 0-3 & 0-3 & LLM \\
Query Satisfaction ($QSAT_u$)   & 1-5 & 1-5 & 1-5 & real users \\
Task Satisfaction ($TSAT_u$)   & 1-5 & 1-5 & 1-5 & real users \\
\bottomrule
\end{tabular}
\end{table}

\subsection{TRUE}
We propose \textbf{TRUE}, a \textit{Task-aware Rubric-based Usefulness Evaluation} framework inspired by SAT/DSAT rubrics for user satisfaction in conversational AI systems~\cite{lin2024interpretable}. User search sessions are inherently personalized and contain diverse signals reflecting relevance, satisfaction, and usefulness. As no standardized method exists for usefulness judgment, we apply TRUE on top of curated baseline and session-based methods (Section~\ref{sec:comparison_methods}) to structure LLM reasoning and improve prediction performance. TRUE follows a five-phase pipeline (see Figure~\ref{fig:true_formulation}): Data Sampling, Reasoning Generation, Rubric Candidate Generation, Rubric Finalization, and Usefulness Label Generation. Details of each phase are described below.

\subsubsection{\textbf{Step 1. User Data Sampling:}}
We begin by sampling user search interactions containing behavioral signals. The THUIR-KDD'19 (hereafter referred to as KDD'19) dataset contains 447 search sessions from 50 participants, comprising 3,046 query--document interactions annotated with usefulness, relevance, satisfaction, and rich behavioral signals such as click-through rate and dwell time (detailed in Section ~\ref{sec:datasets}). We selected KDD'19 for rubric induction because it provides the most comprehensive set of document-level, query-level, task-level, and behavioral features used throughout TRUE (shown in Table~\ref{tab:feature_availability}). Importantly, the rubric is induced only on KDD'19 and subsequently evaluated on the remaining KDD'19 data as well as external datasets, enabling us to assess the generalizability of the induced usefulness criteria.
To ensure generalizability on the KDD'19 dataset as well, we construct a balanced subset $S$ comprising 10\% of the data ($n = 304$), with a uniform distribution over usefulness labels $U_u \in \{0,1,2,3\}$ (Table~\ref{tab:scales_key_measures}). This subset is used for rubric induction and ensures coverage across all usefulness levels. Let $D = \{(q_k, d_k)\}_{k=1}^{N}$ denote the full set of query--document interactions, where each pair $(q_k, d_k)$ corresponds to a clicked document $d_k$ for query $q_k$, annotated with usefulness label $u_k \in \{0,1,2,3\}$. Sampling is performed at the query--document level to enable fine-grained analysis independent of session boundaries. Let $\mathcal{U} = \{0,1,2,3\}$ denote the label set. The dataset label distribution is:

\begin{equation}
p^{*}(u) = \frac{\#\text{interactions with label } u}{\#\text{all interactions}},
\quad u \in \mathcal{U}.
\end{equation}

We construct $S \subset D$ with empirical distribution:

\begin{equation}
\hat{p}(u) = \frac{\#\text{interactions in } S \text{ with label } u}{|S|},
\quad u \in \mathcal{U}.
\end{equation}
The subset is selected such that $\hat{p}(u)$ is balanced across $\{0,1,2,3\}$, using a fixed random seed (42) for reproducibility. This balanced sampling enables the model to better capture relationships between behavioral signals and usefulness.
\subsubsection{\textbf{Step 2. Reasoning Generation:}}
We input the sampled subset $S$ into a reasoning model, GPT-OSS 120B, to generate reasoning patterns for each usefulness label. This step is performed once offline (inference $\sim 30$ minutes for $n=304$) and does not affect runtime efficiency. The model is prompted to explain how implicit and explicit signals, along with document context, correspond to usefulness levels $u \in \{0,1,2,3\}$ across \textbf{task-level} (task satisfaction, task dwell time), \textbf{query-level} (query, query satisfaction, query dwell time), and \textbf{document-level} features (title, summary, alignment, rank, usefulness score, task relevance, query relevance, CTR, URL dwell time, average URL dwell time, and labels: \textit{HR-HU, HR-LU, LR-HU, LR-LU}) (see Table~\ref{tab:feature_availability} and Section~\ref{sec:datasets}). We categorize the relationship between relevance and usefulness across four cases: HR-HU (high relevance--high usefulness), HR-LU (high relevance--low usefulness), LR-HU (low relevance--high usefulness), and LR-LU (low relevance--low usefulness). The model also accounts for exceptional cases such as \textit{HR-LU} (high relevance--low usefulness) and \textit{LR-HU} (low relevance--high usefulness), where relevance and usefulness diverge and move in opposite directions. This step aims to uncover consistent behavioral patterns and produce interpretable rationales for rubric construction (Figure~\ref{fig:true_formulation}). Let $f_{\mathrm{oss}}$ denote the GPT-OSS 120B reasoning model. For sampled data $S$, the model receives:
\[
x_i = \mathrm{Prompt}(S, \mathcal{U})
\]
where $\mathrm{Prompt}(\cdot)$ includes sampled interactions, behavioral signals, document metadata, and usefulness labels across task, query, and document levels, with $\mathcal{U} = \{0,1,2,3\}$. The model generates:
\[
g_i = f_{\mathrm{oss}}(x_i)
\]
where $g_i$ is a textual explanation describing relationships between signals and usefulness labels. The final set of generated rationales is:
\[
\mathcal{G} = \{ g_1, \ldots, g_{|S|} \}.
\]

\subsubsection{\textbf{Step 3. Candidate Rubric Generation:}} We transform the generated reasoning set $\mathcal{G}$ into structured rule candidates using GPT-OSS 120B. The objective of this step is not to learn dataset-specific rules, but to extract \textit{generalizable behavioral and semantic patterns} that characterize usefulness across search sessions.

We feed the generated reasoning set $\mathcal{G}$ back into the GPT-OSS 120B model using a rubric extraction prompt, producing $T=3$ independent candidate rubrics:
\[
r_i \;=\; f_{\mathrm{oss}}^{\mathrm{rubric}}(g_i)
\]
Each candidate rubric represents a distinct abstraction of the reasoning patterns, organizing them into structured rules across usefulness levels $u \in {0,1,2,3}$. All candidates\footnote{Rubric candidate samples, all TRUE variant prompts and codes are available at: \url{https://github.com/moulydewan/LLMUsefulnessLabels.git}} follow a consistent structure across five feature groups: \textit{Relevance, Satisfaction, User Behavior, Title Summary Alignment, and Exceptional Cases} while differing in how these signals are interpreted and combined. This multi-candidate design introduces diversity and improves coverage, enabling the identification of consistent patterns across independently generated rule sets.

\subsubsection{\textbf{Step 4. Rubric Finalization:}}
\begin{figure}[t]
    \centering
    \includegraphics[width=\linewidth]{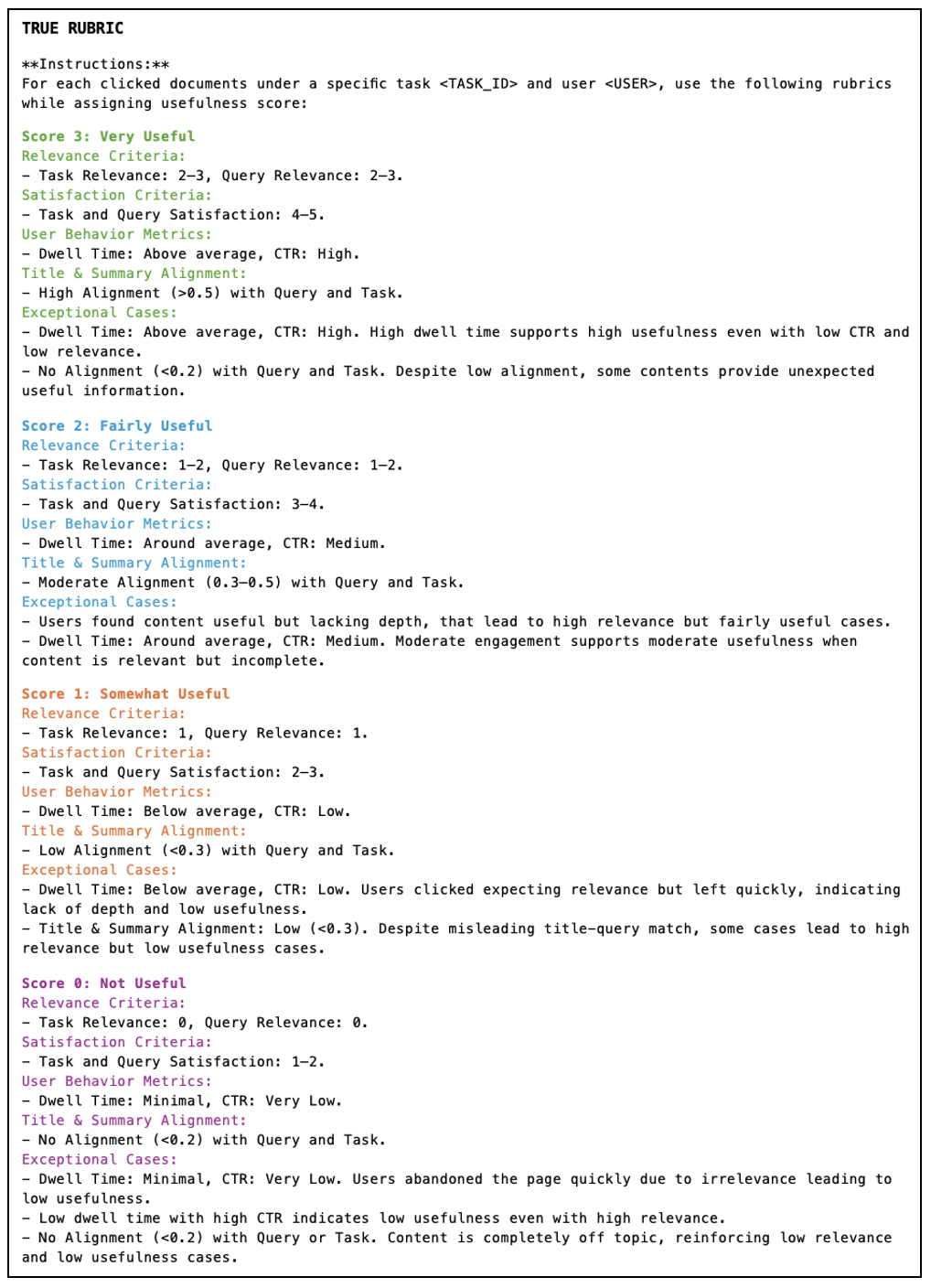}
    \caption{TRUE Rubric used as a prompt.}
    \label{fig:prompt}
\end{figure}

We perform structured manual consolidation over the $T=3$ candidate rubrics to obtain a unified rubric $\mathbf{R}$. This step represents the only point of human intervention in the pipeline. The consolidation focuses on (i) merging overlapping or redundant rules, (ii) resolving inconsistencies in signal interpretation, and (iii) expressing rules in clear, high-level language. Importantly, this step does not introduce new signals or dataset-specific conditions; instead, it produces a \textit{consensus} across candidate rubrics, emphasizing feature groups such as relevance, satisfaction, alignment, user behavior, and exceptional cases. The thresholds are not manually tuned but emerge consistently across LLM-generated rubrics and are consolidated into a unified framework. Human intervention is limited to improving interpretability and coherence, without altering the underlying behavioral relationships or tailoring rules to specific datasets. We further validate this design through a rubric ablation study (Section~\ref{sec:rubricablation}). To quantify manual intervention, we analyze the consolidation of the $T=3$ candidate rubrics into the final unified rubric. Each candidate rubric contains rules across four usefulness levels and multiple feature groups (approximately 16--20 rule blocks per rubric). Approximately 40--50\% of rules remain structurally unchanged, while 50--60\% are rephrased or merged for clarity and redundancy removal, and a small fraction ($\sim$5\%) are removed. The only structural adjustment is the removal of satisfaction signals from exceptional cases to align with our controlled evaluation setting (RQ3). Overall, manual intervention is limited to improving coherence and generalizability, without altering the underlying decision structure. The TRUE rubric adopts a hybrid representation to balance precision and transferability. Signals with well-defined ordinal scales, such as relevance (0--3) and satisfaction (1--5), are expressed using dataset-aligned numeric ranges, while behavioral signals (dwell time, CTR), which vary across datasets, are represented using relative qualitative descriptors (above average, low) to avoid overfitting. Satisfaction signals are treated as supporting contextual features rather than primary decision criteria and are excluded from exceptional cases to mitigate potential bias. The final rubric $\mathbf{R}$ (Figure~\ref{fig:prompt}) defines usefulness across five feature groups using relative, interpretable rules rather than fixed thresholds. Although induced on KDD'19, this design enables direct application to other datasets (QRef and SIGIR16) with minimal normalization. In practice, transfer required only aligning ordinal label scales (mapping relevance from 0--4 in SIGIR16 to 0--3 in KDD'19) and accommodating minor feature differences, while preserving the core decision structure. This ensures generalization without introducing dataset-specific rules.

\begin{table}[t]
\small
\centering
\caption{Feature availability and signal type across datasets.}
\label{tab:feature_availability}
\begin{tabular}{l l c c c}
\toprule
\textbf{Feature Category} & \textbf{Signal} & \textbf{KDD'19} & \textbf{QRef} & \textbf{SIGIR16} \\
\midrule
User identifier & Metadata & \checkmark & \checkmark & \checkmark \\
Session / task identifier & Metadata & \checkmark & \checkmark & \checkmark \\

\multicolumn{5}{l}{\textbf{Task-level Features}} \\
Task description & Explicit & \checkmark & -- & \checkmark \\
Task satisfaction ($TSAT_u$) & Explicit & \checkmark & \checkmark & \checkmark \\
Task dwell time (ms) & Implicit & \checkmark & \checkmark & -- \\
Task relevance score & Explicit & \checkmark & -- & -- \\

\multicolumn{5}{l}{\textbf{Query-level Features}} \\
Initial query & Explicit & -- & \checkmark & -- \\
Query & Explicit & \checkmark & \checkmark & \checkmark \\
Query position in session & Implicit & \checkmark & \checkmark & \checkmark \\
Query satisfaction ($QSAT_u$) & Explicit & \checkmark & \checkmark & \checkmark \\
Query dwell time (ms) & Implicit & \checkmark & \checkmark & -- \\
Query relevance ($R_a$)+ & Explicit & \checkmark & -- & \checkmark \\

\multicolumn{5}{l}{\textbf{Document-level Features}} \\
Document URL & Metadata & \checkmark & \checkmark & \checkmark \\
Document title & Explicit & \checkmark & \checkmark & \checkmark \\
Document snippet & Explicit & \checkmark & \checkmark & \checkmark \\
Alignment Label & Implicit & \checkmark & \checkmark & \checkmark \\
Rank of document & Implicit & \checkmark & \checkmark & \checkmark \\
Usefulness ($U_u$)* & Explicit & \checkmark & \checkmark & \checkmark \\
URL dwell time (ms) & Implicit & \checkmark & -- & \checkmark \\
Avg. URL dwell time (ms) & Implicit & \checkmark & \checkmark & \checkmark \\
Click-through rate (CTR) & Implicit & \checkmark & \checkmark & -- \\
Label (High/Low - Rel./Use.) & Implicit & \checkmark & -- & \checkmark \\

\bottomrule
\end{tabular}
\end{table}

\subsubsection{\textbf{Step 5. Usefulness Generation with TRUE:}}
Finally, we apply the \textsc{TRUE} rubrics to generate usefulness labels, $U_{\text{llm}}$, using multiple LLMs (Figure~\ref{fig:true_formulation}). TRUE defines rubric rules across five feature groups: relevance, satisfaction, user behavior, title--summary alignment, and exceptional cases. For each usefulness label $u \in \{0,1,2,3\}$, the framework provides a corresponding set of rules within each group. We apply these rules on top of baseline and session-level representations, resulting in four evaluation settings: TRUE\textsubscript{Baseline}, TRUE\textsubscript{Session}, TRUE\textsubscript{Rubric+Baseline}, and TRUE\textsubscript{Rubric+Session}. The first two operate without rubric constraints, while the latter two explicitly enforce the induced TRUE rules. Usefulness labels $U_{\text{llm}}$ are generated under all settings using a consistent evaluation framework. Details of baseline and session representations are provided in Section~\ref{sec:comparison_methods}. 
\subsubsection{\textbf{Data Leakage and Circularity.}}
To mitigate data leakage and circularity concerns in LLM-based evaluation \citep{dietz2025principles}, we explicitly separate rubric induction from application. The TRUE rubric is constructed on a held-out 10\% subset of KDD'19, while usefulness label generation is performed exclusively on the remaining 90\%. We further reduce leakage by using different models at each stage: rubric induction (reasoning and construction) is performed with GPT-OSS 120B, while usefulness inference is conducted using separate LLMs (LLaMA and GPT variants). This prevents the same model from both generating and applying the rubric, limiting implicit memorization. The rubric itself is defined as high-level, interpretable rules over observable signals (relevance, satisfaction, user behavior, alignment and exceptional cases), rather than relying on direct access to ground-truth usefulness labels at inference time. Although labels are used during rubric induction, they serve only to extract generalizable patterns rather than train a predictive model. To assess robustness, the same rubric is applied without modification to additional datasets (QRef and SIGIR16), with only minimal normalization to align label scales and feature differences. This demonstrates that the induced rules capture generalizable signals rather than dataset-specific artifacts. 

We acknowledge that residual circularity may remain due to the use of labeled data during rubric induction. However, we treat the rubric as an interpretable abstraction of observed patterns, not a learned model. Importantly, comparison methods (baseline and session-based approaches) are entirely label-independent, ensuring that evaluation is not solely driven by label-informed components.

\begin{table*}[t]
    \caption{Statistics of behavior logs and rubric splits. TRUE rubric is induced only on a 10\% subset of KDD'19 ($n=304*$).}
    \small
    \label{tab:behavior_logs}
    \centering
    \begin{tabular}{lcccccccccc}
        \toprule
        \textbf{Dataset} & \textbf{\#Tasks} & \textbf{\#Users} & \textbf{\#Sessions} & \textbf{\#Qry} & \textbf{\#Ck. Docs} & \textbf{\#Ck. Doc / Qry} & \textbf{\#Rows} & \textbf{\#Held out (10\%)} & \textbf{\#Eval (90\%)} & \textbf{\#Difficulty} \\
        \midrule
        KDD'19 & 9 & 50 & 447 & 734 & 1,430 & 1.95 & 3,046 & 304* & 2,742 & Complex \\
        QRef         & -- & 42 & 2,024 & 4,555 & 7,126 & 1.56 & 7,380 & 736 & 6,644 & Normal \\
        SIGIR16      & 15 & 25 & 225 & 367 & 657 & 1.79 & 1,507 & 148 & 1,359 & Normal \\
        \bottomrule
    \end{tabular}
\end{table*}

\section{Datasets and Features} 
\label{sec:datasets}

Our objective is to evaluate the ability of LLMs to generate usefulness labels using implicit and explicit user behavior signals \citep{chen2017meta}, along with relevance scores and document-level features. We use three real-world search log datasets: THUIR-KDD'19 \citep{mao2016does}, TianGong-QRef \citep{chen2021towards}, and SIGIR16 \citep{10.1145/3077136.3080750, mao2016does} (Table~\ref{tab:behavior_logs}). All datasets include user behavior signals, document metadata, and human-annotated usefulness labels (Table~\ref{tab:feature_availability}), while KDD'19 and SIGIR16 additionally provide relevance labels (Table~\ref{tab:scales_key_measures}). From session logs, we derive online behavior features such as click-through rate (CTR), URL dwell time, query dwell time, task dwell time, and average URL dwell time per user (availability varies by dataset). KDD'19 and SIGIR16 contain longer, task-oriented sessions with multiple queries per task, whereas QRef consists of shorter, open-ended sessions with a larger number of sessions (2,024). For KDD'19 and SIGIR16, usefulness (U) labels range from 0 to 3; we group them into Low (0--1) and High (2--3), forming four categories: High Relevance–High Usefulness, High Relevance–Low Usefulness, Low Relevance–High Usefulness, and Low Relevance–Low Usefulness. We also compute title–summary alignment using cosine similarity (all-MiniLM-L6-v2) between query/task and document title/summary, as High (>0.5), Moderate (0.3–0.5), Low (<0.3), and No Alignment (<0.2).

Additionally, all datasets provide explicit \textit{query-level} and \textit{task-level satisfaction} annotations. We treat these signals as auxiliary features rather than direct predictors of usefulness. While satisfaction is typically considered a downstream outcome of usefulness, we include it to capture aggregated user feedback not fully observable through interaction features alone. Our goal is not to model the causal relationship between satisfaction and usefulness, but to evaluate LLMs under a full-information setting. Specifically, we assess whether LLMs can infer usefulness when provided with all available signals, representing an upper-bound, answer-level evaluation of their ability to integrate heterogeneous signals. We further analyze the impact of satisfaction through controlled experiments (Section~\ref{sec:rq3_satisfaction}), comparing settings with and without satisfaction.

\section{Methodology}
Our methodology evaluates usefulness prediction under four settings: \textbf{baseline}, \textbf{session}, and their rubric-enhanced variants using \textbf{TRUE}. We first aggregate implicit and explicit user behavior signals across users and tasks to capture search interactions, and transform them into structured inputs for LLMs. TRUE is then applied to both baseline and session representations, yielding four methods: \textsc{TRUE}\textsubscript{Baseline}, \textsc{TRUE}\textsubscript{Session}, \textsc{TRUE}\textsubscript{Rubric+Baseline}, and \textsc{TRUE}\textsubscript{Rubric+Session}. All methods are evaluated across five LLMs to generate usefulness scores for retrieved documents.

\subsection{Prompt Techniques}
To control for prompt sensitivity~\citep{liu2023pre}, we standardize all experiments using a zero-shot DNA prompt~\citep{thomas_large_2024}. The DNA template provides a consistent structure with descriptive, narrative, and aspects components, enabling structured reasoning over search context. In the descriptive section, the LLM is instructed to act as a search quality rater following Google’s guidelines~\citep{Google2022}, with all input signals (relevance, satisfaction, and behavioral metrics) explicitly defined. The aspects section employs a step-by-step Chain-of-Thought strategy~\citep{wei2022chain} to decompose the labeling task and capture relationships across queries, tasks, and interactions. We evaluate four variants under this shared prompt: \textit{baseline} (no session context or rubric), \textit{session} (with session context), \textit{baseline+rubric}, and \textit{session+rubric}. This design ensures that all methods share a common prompting foundation, with differences arising only from the inclusion of session context and rubric constraints.

\subsection{Comparison Methods}
This section presents our curated baseline methods and reviews recent work on LLM-based usefulness judgment used for comparison in our evaluation.
\label{sec:comparison_methods}
\vspace{-6pt}
\subsubsection{Baseline Data}\label{subsubsection:baseline data}
We define the baseline by treating each query--document interaction as an independent instance, without incorporating session context. Each instance includes a user query, a clicked document, and associated behavioral and relevance signals. This representation serves as the reference point for evaluating all methods, reflecting a simple and generalizable formulation for LLM-based usefulness prediction. The KDD'19, QRef, and SIGIR16 datasets contain approximately 3,046, 7,380, and 1,507 instances, respectively (Table~\ref{tab:behavior_logs}).
\vspace{-6pt}
\subsubsection{Session Data}
In this setting, we group all queries and their associated clicked documents within a user’s task into a single batch, referred to as \textbf{session data}. This representation preserves session-level context, reflecting how users iteratively refine queries to complete a task. Each batch corresponds to one user task session and is provided to the LLM as a unified input. This results in approximately 447 sessions for KDD'19, 2,024 for QRef, and 225 for SIGIR16 (Table~\ref{tab:behavior_logs}).
\vspace{-6pt}
\subsubsection{State-of-the-Art Comparison Methods}
To evaluate TRUE, we compare against both traditional machine learning and LLM-based usefulness prediction methods. For traditional baselines, we include Linear Regression as a simple feature-aggregation model, along with the Gradient Boosting Decision Tree (GBDT) model of Mao et al.~\citep{mao_understanding_2017}, a strong benchmark that leverages contextual, content, and behavioral features such as dwell time and CTR, but does not incorporate explicit relevance or satisfaction signals. For LLM-based approaches, we include CLUE~\citep{wang2025clue}, which performs usefulness judgment on SIGIR16 and KDD'19, enabling direct comparison with our method TRUE. In addition, we evaluate four internal variants: \textsc{TRUE}\textsubscript{Baseline}, \textsc{TRUE}\textsubscript{Session}, \textsc{TRUE}\textsubscript{Rubric+Baseline}, and \textsc{TRUE}\textsubscript{Rubric+Session}, isolating the effects of session context and rubric constraints.

\begin{table*}[t]
\caption{Spearman's rank correlation comparison across models and methods on KDD'19, QRef, and SIGIR16 datasets. \textbf{Bold} indicates methods that outperform the corresponding Baseline within each row. * denotes the three highest Spearman correlations observed in SIGIR16. Statistical significance was assessed using paired permutation tests on per-task and per-query Spearman correlations, with Holm--Bonferroni correction for multiple comparisons. All improvements over the Baseline are statistically significant ($p < 0.05$).}
\label{tab:spearman_results}
\centering
\small
\setlength{\tabcolsep}{4pt}
\renewcommand{\arraystretch}{1.15}
\begin{tabular}{llccc ccc ccc}
\toprule
\textbf{LLM} & \textbf{Method}
& \multicolumn{3}{c}{\textbf{KDD'19}}
& \multicolumn{3}{c}{\textbf{QRef}}
& \multicolumn{3}{c}{\textbf{SIGIR16}} \\
\cmidrule(lr){3-5} \cmidrule(lr){6-8} \cmidrule(lr){9-11}
 &  
 & \textbf{Overall} & \textbf{Task} & \textbf{Query}
 & \textbf{Overall} & \textbf{Task} & \textbf{Query}
 & \textbf{Overall} & \textbf{Task} & \textbf{Query} \\
\midrule
\multirow{4}{*}{LLaMA 3.3 70B}
 & \textsc{TRUE}\textsubscript{Baseline} & 0.41 & 0.34 & 0.27 & 0.38 & 0.40 & 0.38 & 0.34 & 0.30 & 0.21 \\
 & \textsc{TRUE}\textsubscript{Session} & 0.36 & 0.31 & \textbf{0.30} & 0.37 & \textbf{0.41} & \textbf{0.46} & \textbf{0.50} & \textbf{0.49} & \textbf{0.52} \\
 & \textsc{TRUE}\textsubscript{Rubric+Baseline} & \textbf{0.46} & \textbf{0.40} & 0.26 & 0.36 & 0.36 & 0.11 & 0.26 & 0.22 & 0.06 \\
 & \textsc{TRUE}\textsubscript{Rubric+Session} & \textbf{0.41} & \textbf{0.37} & \textbf{0.32} & 0.37 & \textbf{0.43} & \textbf{0.47} & \textbf{0.53*} & \textbf{0.51} & \textbf{0.57*} \\
\midrule

\multirow{4}{*}{LLaMA 3.2 3B}
 & \textsc{TRUE}\textsubscript{Baseline} & 0.23 & 0.13 & 0.03 & 0.13 & 0.20 & 0.22 & 0.12 & 0.09 & 0.11 \\
 & \textsc{TRUE}\textsubscript{Session} & \textbf{0.27} & \textbf{0.23} & \textbf{0.21} & \textbf{0.30} & \textbf{0.37} & \textbf{0.47} & \textbf{0.23} & \textbf{0.23} & \textbf{0.24} \\
 & \textsc{TRUE}\textsubscript{Rubric+Baseline} & \textbf{0.31} & \textbf{0.23} & \textbf{0.17} & \textbf{0.27} & \textbf{0.24} & 0.19 & 0.10 & \textbf{0.09} & \textbf{0.11} \\
 & \textsc{TRUE}\textsubscript{Rubric+Session} & \textbf{0.30} & \textbf{0.27} & \textbf{0.20} & \textbf{0.30} & \textbf{0.38} & \textbf{0.49} & \textbf{0.27} & \textbf{0.27} & \textbf{0.27} \\
\midrule

\multirow{4}{*}{LLaMA 3.1 8B}
 & \textsc{TRUE}\textsubscript{Baseline} & 0.30 & 0.25 & 0.15 & 0.32 & 0.36 & 0.39 & 0.29 & 0.26 & 0.23 \\
 & \textsc{TRUE}\textsubscript{Session} & \textbf{0.33} & \textbf{0.29} & \textbf{0.25} & 0.26 & 0.35 & \textbf{0.46} & \textbf{0.36} & \textbf{0.35} & \textbf{0.33} \\
 & \textsc{TRUE}\textsubscript{Rubric+Baseline} & \textbf{0.39} & \textbf{0.34} & \textbf{0.21} & \textbf{0.33} & 0.30 & 0.22 & 0.28 & 0.22 & \textbf{0.30} \\
 & \textsc{TRUE}\textsubscript{Rubric+Session} & \textbf{0.38} & \textbf{0.34} & \textbf{0.23} & 0.25 & 0.35 & \textbf{0.48} & \textbf{0.42} & \textbf{0.39} & \textbf{0.43} \\
\midrule

\multirow{4}{*}{GPT-3.5 Turbo}
 & \textsc{TRUE}\textsubscript{Baseline} & 0.32 & 0.27 & 0.23 & 0.33 & 0.26 & 0.07 & 0.27 & 0.23 & 0.12 \\
 & \textsc{TRUE}\textsubscript{Session} & \textbf{0.36} & \textbf{0.31} & \textbf{0.32} & 0.31 & \textbf{0.42} & \textbf{0.46} & \textbf{0.43} & \textbf{0.41} & \textbf{0.49} \\
 & \textsc{TRUE}\textsubscript{Rubric+Baseline} & \textbf{0.38} & \textbf{0.33} & \textbf{0.25} & \textbf{0.36} & \textbf{0.29} & \textbf{0.10} & \textbf{0.32} & \textbf{0.28} & \textbf{0.25} \\
 & \textsc{TRUE}\textsubscript{Rubric+Session} & \textbf{0.37} & \textbf{0.34} & \textbf{0.32} & \textbf{0.39} & \textbf{0.29} & \textbf{0.46} & \textbf{0.50} & \textbf{0.49} & \textbf{0.60*} \\
\midrule

\multirow{4}{*}{GPT-4o Mini}
 & \textsc{TRUE}\textsubscript{Baseline} & 0.39 & 0.32 & 0.28 & 0.31 & 0.33 & 0.35 & 0.38 & 0.34 & 0.37 \\
 & \textsc{TRUE}\textsubscript{Session} & 0.33 & 0.29 & 0.23 & \textbf{0.38} & \textbf{0.42} & \textbf{0.51} & \textbf{0.42} & \textbf{0.40} & \textbf{0.41} \\
 & \textsc{TRUE}\textsubscript{Rubric+Baseline} & \textbf{0.43} & \textbf{0.36} & 0.27 & \textbf{0.37} & \textbf{0.34} & 0.33 & \textbf{0.40} & \textbf{0.37} & \textbf{0.40} \\
 & \textsc{TRUE}\textsubscript{Rubric+Session} & 0.35 & 0.31 & 0.23 & \textbf{0.41} & \textbf{0.40} & \textbf{0.42} & \textbf{0.41} & \textbf{0.38} & \textbf{0.37} \\

\bottomrule
\end{tabular}
\end{table*}

\section{Experimental Setup}
In this section, we describe our experimental setup and evaluated LLMs. All experiments use zero-shot prompting based on the DNA template, where LLMs assign usefulness labels to clicked documents as: \textbf{3 = Very Useful}, \textbf{2 = Fairly Useful}, \textbf{1 = Somewhat Useful}, and \textbf{0 = Not Useful at all}. We evaluate OpenAI’s GPT-3.5 Turbo and GPT-4o Mini~\citep{achiam2023gpt}, along with Meta’s LLaMA models (Llama 3.3 70B-Instruct, Llama 3.2 3B-Instruct, and Llama 3.1 8B-Instruct)~\citep{dubey2024llama}. Experiments are conducted using AWS Bedrock for LLaMA and OpenAI APIs for GPT models, following \cite{thomas_large_2024} with temperature = 0 and top\_p = 1. We include GPT models for their strong performance in relevance labeling while also evaluating competitive open-source alternatives. Ground-truth evaluation uses user-provided usefulness labels ($U_u$). Given the ordinal nature of these labels (0–3), we report Spearman’s rank correlation ($\rho$) for internal comparisons, and additionally report Spearman’s $\rho$, Cohen’s $\kappa$, precision, recall, F1, and MAE for comparisons with state-of-the-art methods.

\section{Results}
Our experiments address four main contributions evaluating the effectiveness of LLM-generated usefulness labels. In the following sections, we discuss each RQ and present the corresponding evaluation methods and results in detail.

\begin{figure*}[t]
    \centering
    \includegraphics[width=0.9\linewidth]{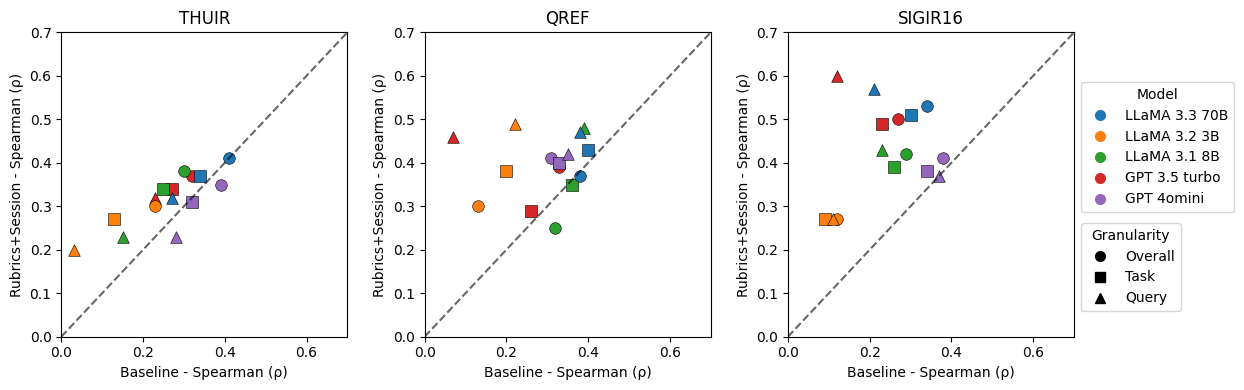}
    \caption{Comparison of Spearman correlation between \textsc{TRUE}\textsubscript{Baseline} and \textsc{TRUE}\textsubscript{Rubric+Session} across datasets and granularities. Points above the diagonal indicate improved alignment with human usefulness judgments.}
    \label{fig:baseline-vs-session}
\end{figure*}

\begin{table}[t]
\centering
\setlength{\tabcolsep}{5pt}
\renewcommand{\arraystretch}{1.15}
\caption{Spearman's rank correlation comparison with SOTA baselines on the SIGIR16 dataset. \textbf{Bold} indicates correlations greater than the corresponding minimum SOTA value within a row, and * denotes the three highest Spearman correlations observed in the dataset. All improvements over the SOTA baseline are statistically significant ($p < 0.05$).}

\label{tab:sigir16-results}
\scriptsize
\begin{tabular}{lcccccc}
\toprule
\textbf{Model} & \textbf{S-$\rho$} & \textbf{C-$\kappa$} & \textbf{Precision} & \textbf{Recall} & \textbf{F1} & \textbf{MAE} \\
\midrule
\textit{$U_a$ (Human–Human)} \citep{wang2025clue} & \textit{0.35} & \textit{0.17} & \textit{0.39} & \textit{0.38} & \textit{0.38} & \textit{0.92} \\
\midrule
\textit{Linear Regression} & \textit{0.30} & \textit{0.02} & \textit{0.36} & \textit{0.27} & \textit{0.17} & \textit{0.99} \\
\textit{GBDT\textsubscript{iid} (SOTA) \citep{mao_understanding_2017}} & \textit{0.42} & \textit{0.24} & \textit{0.40} & \textit{0.44} & \textit{0.44} & \textit{0.86} \\
\textit{GBDT\textsubscript{ood} (SOTA) \citep{mao_understanding_2017}} & \textit{0.31} & \textit{0.16} & \textit{0.40} & \textit{0.40} & \textit{0.37} & \textit{1.01} \\
\textit{CLUE-3.5 (SOTA) \citep{wang2025clue}} & \textit{0.37} & \textit{0.16} & \textit{0.34} & \textit{0.40} & \textit{0.36} & \textit{0.99} \\
\textit{CLUE-4o (SOTA) \citep{wang2025clue}} & \textit{0.38} & \textit{0.18} & \textit{0.38} & \textit{0.41} & \textit{0.38} & \textit{0.98} \\
\midrule
TRUE\textsubscript{Baseline} 3B & 0.12 & 0.01 & 0.29 & 0.26 & 0.12 & 1.69 \\
TRUE\textsubscript{Baseline} 8B & 0.29 & 0.09 & 0.32 & 0.34 & 0.27 & 1.22 \\
TRUE\textsubscript{Baseline} 70B & \textbf{0.34} & 0.10 & \textbf{0.39} & 0.36 & 0.29 & 1.10 \\
TRUE\textsubscript{Baseline} 3.5 & 0.27 & 0.05 & 0.30 & 0.30 & 0.23 & 1.28 \\
TRUE\textsubscript{Baseline} 4o & \textbf{0.38} & 0.05 & 0.32 & 0.31 & 0.22 & 1.28 \\
\midrule
TRUE\textsubscript{Session} 3B & 0.23 & 0.05 & 0.24 & 0.24 & 0.20 & 1.27 \\
TRUE\textsubscript{Session} 8B & \textbf{0.36} & 0.09 & \textbf{0.34} & 0.34 & 0.27 & 1.16 \\
TRUE\textsubscript{Session} 70B & \textbf{0.50*} & 0.13 & \textbf{0.39} & 0.38 & 0.32 & \textbf{0.99} \\
TRUE\textsubscript{Session} 3.5 & \textbf{0.43} & 0.12 & \textbf{0.40} & 0.37 & 0.32 & \textbf{0.94} \\
TRUE\textsubscript{Session} 4o & \textbf{0.42} & 0.09 & \textbf{0.37} & 0.34 & 0.30 & 1.02 \\
\midrule
TRUE\textsubscript{Rubric+Baseline} 3B & 0.10 & 0.01 & 0.31 & 0.26 & 0.09 & 1.77 \\
TRUE\textsubscript{Rubric+Baseline} 8B & 0.28 & 0.10 & 0.32 & 0.34 & 0.30 & 1.03 \\
TRUE\textsubscript{Rubric+Baseline} 70B & 0.26 & 0.11 & \textbf{0.35} & 0.35 & 0.30 & 1.14 \\
TRUE\textsubscript{Rubric+Baseline} 3.5 & \textbf{0.32} & 0.08 & \textbf{0.36} & 0.34 & 0.26 & 1.18 \\
TRUE\textsubscript{Rubric+Baseline} 4o & \textbf{0.40} & 0.05 & 0.31 & 0.31 & 0.20 & 1.31 \\
\midrule
TRUE\textsubscript{Rubric+Session} 3B & 0.27 & 0.06 & 0.25 & 0.25 & 0.19 & 1.32 \\
TRUE\textsubscript{Rubric+Session} 8B & \textbf{0.42} & 0.10 & 0.28 & 0.28 & 0.23 & 1.10 \\
TRUE\textsubscript{Rubric+Session} 70B & \textbf{0.53*} & \textbf{0.19} & \textbf{0.43} & \textbf{0.43} & \textbf{0.38} & \textbf{0.88} \\
TRUE\textsubscript{Rubric+Session} 3.5 & \textbf{0.50*} & 0.14 & \textbf{0.41} & \textbf{0.40} & 0.32 & \textbf{0.98} \\
TRUE\textsubscript{Rubric+Session} 4o & \textbf{0.41} & 0.09 & \textbf{0.37} & 0.34 & 0.28 & 1.11 \\
\bottomrule
\end{tabular}
\end{table}

\subsection{RQ1 \& RQ2: Performance of TRUE}
\subsubsection{Comparison with TRUE variants (Session and Rubric integration):}
We evaluate performance across three datasets: KDD'19 (longer sessions), QRef (shorter sessions), and SIGIR16 (longer sessions) using four internal methods. Due to limited prior work, we construct a baseline prompt as a reference and compare \textsc{TRUE}\textsubscript{Session}, \textsc{TRUE}\textsubscript{Rubric+Baseline}, and \textsc{TRUE}\textsubscript{Rubric+Session} against \textsc{TRUE}\textsubscript{Baseline}. We further compare TRUE with traditional machine learning baselines, including Linear Regression and GBDT, as well as the recent LLM-based method CLUE. Performance is evaluated at three granularities: overall, task, and query. Overall correlation is computed across all interactions, while task- and query-level correlations are computed per group and then averaged to capture differences in granularity. As shown in Table~\ref{tab:spearman_results}, on KDD'19, \textsc{TRUE}\textsubscript{Rubric+Baseline} and \textsc{TRUE}\textsubscript{Rubric+Session} consistently outperform \textsc{TRUE}\textsubscript{Baseline} across most levels, with the highest correlation (0.46) achieved by LLaMA 3.3 70B using the rubric method. Session-aware and rubric-based methods particularly benefit smaller models (LLaMA 3.2 3B, LLaMA 3.1 8B, and GPT-3.5 Turbo), improving correlation over the baseline. Since KDD'19 contains longer sessions, these results indicate that incorporating session context and rubric-based reasoning substantially improves correlation compared to the baseline. On QRef, \textsc{TRUE}\textsubscript{Session} consistently outperforms the baseline across all models, with notable gains at the query level. This improvement stems from QRef’s fine-grained query-level granularity and short but higher number of sessions. \textsc{TRUE}\textsubscript{Rubric+Baseline} and \textsc{TRUE}\textsubscript{Rubric+Session}, improve over baseline, especially for LLaMA 3.2 3B, GPT-3.5 Turbo and GPT-4o Mini. However, in LLaMA 3.1 8B the session-aware and rubric methods do not offer improvements. 

For SIGIR16 (longer session), \textsc{TRUE}\textsubscript{Session}, \textsc{TRUE}\textsubscript{Rubric+Baseline}, and \textsc{TRUE}\textsubscript{Rubric+Session} shows consistent correlation improvements compared to \textsc{TRUE}\textsubscript{Baseline} for LLaMA 3.2 3B, GPT-3.5 Turbo and GPT-4o Mini. Moreover, the three strongest Spearman correlations were observed in this dataset on \textsc{TRUE}\textsubscript{Rubric+Session} which is marked with a * in the table. We assess statistical significance using paired permutation tests on per-task and per-query Spearman correlations between model predictions and human usefulness labels. All comparisons are paired against the baseline model and corrected using Holm--Bonferroni to account for multiple comparisons. Statistical significance is reported at $p < 0.05$, and all reported improvements are statistically significant. Overall, session-aware and rubric-based method evaluation improve performances across all three datasets. Therefore, we can state that LLMs can generate reliable and moderate to highly correlated usefulness judgments when provided with representations of search sessions, and a rubric-driven evaluation framework further enhances LLM-based usefulness assessment by yielding consistent improvements over baseline prompting as illustrated in Figure~\ref{fig:baseline-vs-session}. This figure shows that across KDD'19, QRef, and SIGIR16, most models and granularities lie above the diagonal, indicating that \textsc{TRUE}\textsubscript{Rubric+Session} consistently achieves higher Spearman correlation with human judgments than \textsc{TRUE}\textsubscript{Baseline}. 

\subsubsection{Comparison with SOTA baselines:}
Additionally in Table~\ref{tab:sigir16-results}, we include $U_a$ (Human--Human) from prior work \citep{wang2025clue}, which measures agreement between user-provided usefulness judgments and third-party human annotations. This serves as a reference point for the inherent subjectivity of usefulness, with a Spearman’s $\rho$ of 0.35 indicating moderate agreement between humans. Notably, several of our \textsc{TRUE} variants, particularly \textsc{TRUE}\textsubscript{Rubric+Session} (70B), achieve substantially higher correlations (up to 0.53), suggesting that LLM-based usefulness prediction can operate at or beyond the level of observed human agreement. We further compare against strong supervised baselines, a Linear Regression model trained on user behavior features (task dwell time, query dwell time, query position, rank, URL dwell time, average URL dwell time), following prior work \citep{mao_understanding_2017}, achieves $\rho = 0.30$, indicating that simple feature aggregation provides a reasonable but limited approximation of usefulness.  Stronger models such as GBDT\textsubscript{iid} and GBDT\textsubscript{ood} incorporate a richer feature set spanning user behavior (\textit{doc\_click\_order, doc\_dwell\_time, session\_end}), content (\textit{doc\_content\_text, query\_string\_text, task description\_text}), and contextual signals (\textit{query\_total\_click\_number, query\_clicked\_ranks\_list, query\_max\_clicked\_rank, avg\_doc\_dwell time\_in\_query}), achieving improved performance ($\rho = 0.42$) \citep{wang2025clue}. While the exact feature sets differ across models due to implementation constraints in prior work, we align them as closely as possible using overlapping behavioral, content, and contextual signals. This ensures a reasonable comparison while reflecting realistic modeling differences between feature-based and LLM-based approaches. Despite leveraging explicit supervised learning and richer feature engineering, these models are consistently outperformed by our \textsc{TRUE} variants in Table~\ref{tab:sigir16-results}, which do not require task-specific training. This demonstrates the effectiveness of rubric-guided reasoning in capturing complex usefulness signals beyond traditional feature-based approaches.

\label{sec:rq3_satisfaction}
\begin{figure}[t]
    \centering
    \includegraphics[width=\linewidth]{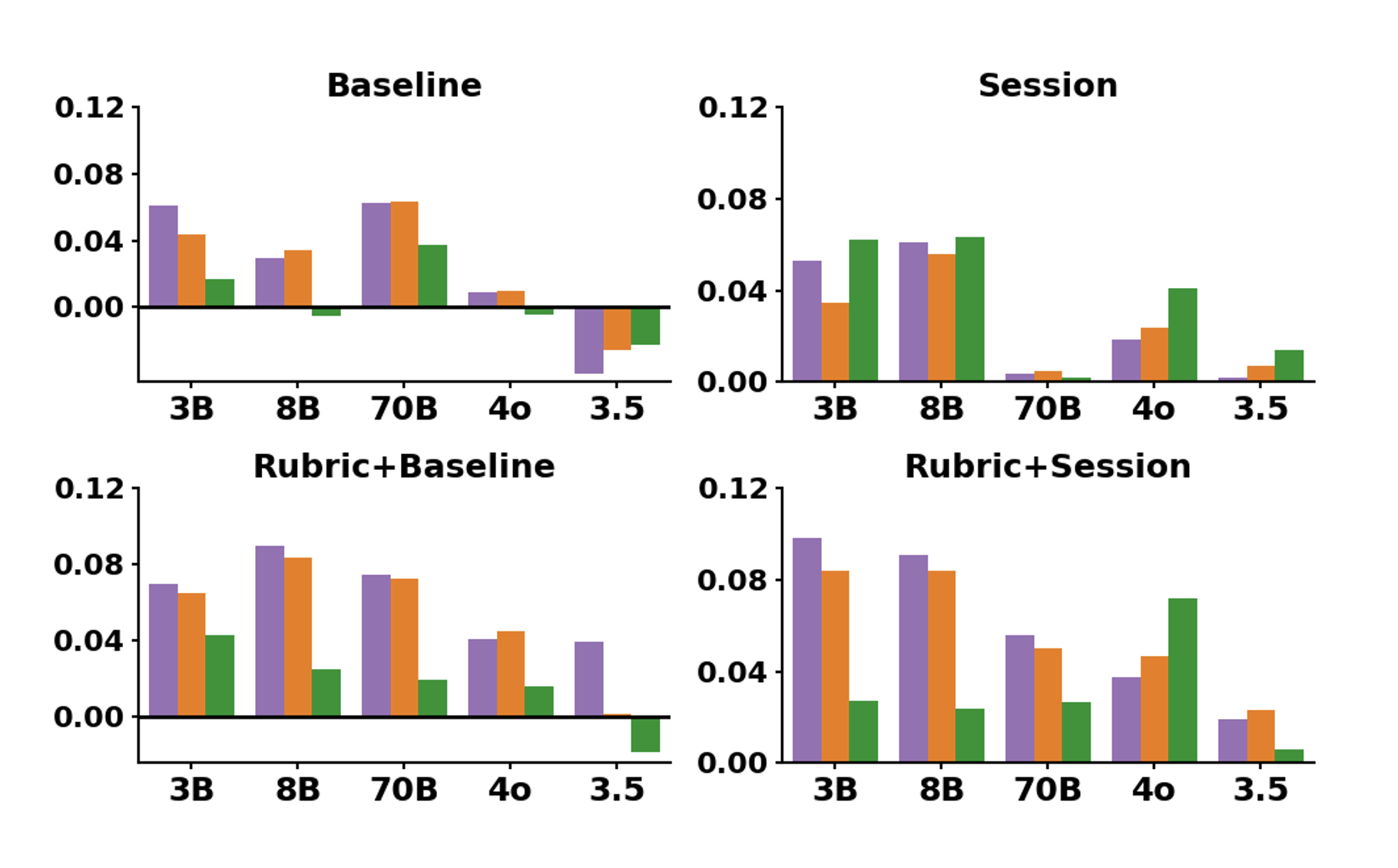}
    \caption{($\Delta$ Spearman = Sat $-$ No Sat)  across models and levels: overall, task, query in KDD'19 dataset.}
    \label{fig:satisfaction}
\end{figure}

\begin{table}[t]
\centering
\small
\setlength{\tabcolsep}{3pt}
\begin{tabular}{llcccccccc}
\toprule
 &  & \multicolumn{2}{c}{B} & \multicolumn{2}{c}{S} & \multicolumn{2}{c}{R+B} & \multicolumn{2}{c}{R+S} \\
\cmidrule(lr){3-4} \cmidrule(lr){5-6} \cmidrule(lr){7-8} \cmidrule(lr){9-10}
Model & Level & $\Delta$ & $p$ & $\Delta$ & $p$ & $\Delta$ & $p$ & $\Delta$ & $p$ \\
\midrule

LLaMA 3.2 3B & Task & 0.04 & 0.14 & 0.04 & 0.70 & 0.07 & 0.08 & 0.08 & \textbf{0.02} \\
     & Query & 0.02 & 0.57 & 0.06 & 0.61 & 0.04 & 0.18 & 0.03 & 0.48 \\

LLaMA 3.1 8B & Task & 0.04 & \textbf{0.01} & 0.06 & \textbf{0.01} & 0.08 & \textbf{0.00} & 0.08 & \textbf{0.00} \\
     & Query & -0.01 & 0.77 & 0.06 & \textbf{0.00} & 0.03 & 0.19 & 0.02 & 0.22 \\

LLaMA 3.3 70B & Task & 0.06 & \textbf{0.00} & 0.01 & 0.66 & 0.07 & \textbf{0.00} & 0.05 & \textbf{0.00} \\
     & Query & 0.04 & \textbf{0.00} & 0.00 & 0.84 & 0.02 & 0.40 & 0.03 & 0.07 \\

GPT 4omini & Task & 0.01 & 0.58 & 0.02 & \textbf{0.05} & 0.05 & \textbf{0.02} & 0.05 & \textbf{0.01} \\
     & Query & -0.00 & 0.82 & 0.04 & \textbf{0.03} & 0.02 & 0.46 & 0.07 & \textbf{0.00} \\

GPT 3.5  & Task & -0.03 & 0.11 & 0.01 & 0.39 & 0.00 & 0.96 & 0.02 & 0.21 \\
     & Query & -0.02 & 0.44 & 0.01 & 0.22 & -0.02 & 0.40 & 0.01 & 0.64 \\

\midrule
All models  & Overall & 0.03 & 0.26 & 0.03 & 0.09 & 0.06 & \textbf{0.00} & 0.06 & \textbf{0.02} \\

\bottomrule
\end{tabular}
\caption{Effect of satisfaction on usefulness prediction ($\Delta$ = Sat $-$ No-Sat) in KDD'19 dataset. Statistically significant results ($p < 0.05$) are bold.}
\label{tab:satisfaction_effect}
\end{table}
\subsection{RQ3: Effect of Satisfaction}
We analyze the impact of incorporating satisfaction signals on usefulness prediction by comparing performance with and without satisfaction across models and methods (Table~\ref{tab:satisfaction_effect}, Figure~\ref{fig:satisfaction}) in KDD'19. Overall, satisfaction yields limited and inconsistent improvements in baseline and session-based settings, with small effect sizes and largely insignificant gains. In contrast, when combined with TRUE, satisfaction leads to substantial improvements across multiple models, with average gains of $\Delta \approx 0.06$ in Spearman correlation. This indicates that satisfaction is not a reliable standalone predictor of usefulness, but can serve as an auxiliary signal. We further observe that the impact of satisfaction is more pronounced at the task level than at the query level. This suggests that satisfaction primarily operates as an aggregate signal reflecting overall task success rather than fine-grained query-level usefulness. Notably, the minimal insignificant improvements at the query level indicate that incorporating query-level satisfaction signals does not affect usefulness prediction, implying that such signals may be optional in fine-grained settings. Additionally, the effectiveness of satisfaction varies across models, with mid-to-large models (e.g., LLaMA 3.1 8B, 70B, GPT-4o Mini) benefiting more consistently, while smaller or weaker models exhibit negligible or unstable effects. Taken together, these results highlight that the utility of satisfaction depends critically on both the modeling framework and the level of granularity at which it is applied. Importantly, while satisfaction can provide measurable improvements, its impact is not sufficient to fundamentally change usefulness judgments, instead acting as a secondary signal. Given that our evaluation is conducted at the answer level, these results indicate that usefulness can be reliably inferred even in the absence of explicit satisfaction signals.

\subsection{RQ4: Feature Ablation Study}
We conduct an ablation study to identify the most critical signals for LLM-based usefulness prediction. Features are grouped into query (Q), document (D), relevance (R), satisfaction (S), and user behavior (U) (Table~\ref{tab:ablation}). Query and document features are held constant, while R, S, and U are varied to isolate their contributions, reflecting practical constraints where prompt length and cost are limited. We evaluate LLaMA 3.3 70B-Instruct and GPT-4o Mini under the \textsc{TRUE}\textsubscript{Rubric+Session} setting on KDD'19. Combining all feature groups (R+S+U) yields the best performance ($\rho = 0.41$ and $0.35$), indicating that usefulness is best captured through a combination of semantic relevance, user feedback, and behavioral signals. Using only relevance and satisfaction (R+S) achieves comparable performance ($\rho = 0.39$ and $0.35$), suggesting that these two signals account for most of the predictive power. In contrast, user behavior alone performs substantially worse ($\rho = 0.30$ and $0.28$), indicating that behavioral signals are insufficient without semantic grounding. Satisfaction alone also provides limited gains, reinforcing its role as a complementary signal. Overall, relevance emerges as the primary signal, with satisfaction providing additional context and user behavior acting as a supporting factor. These results highlight that effective usefulness prediction requires a balanced combination of signals, while also suggesting that selective feature inclusion can reduce prompt cost without significant performance loss.
\begin{table}[t]
\caption{Ablation study of features in best performing LLaMA 3.3 70B and GPT-4o Mini with \textsc{TRUE}\textsubscript{Rubric+Session} data on KDD'19. Bold indicates the best performing across both models.}
\label{tab:ablation}
\small
\centering
\setlength{\tabcolsep}{6pt}           % column padding
\renewcommand{\arraystretch}{0.95}    % row height
\begin{tabular*}{\columnwidth}{@{\extracolsep{\fill}} l c c}
\toprule
\textbf{Features} & \textbf{LLaMA 3.3 70B} & \textbf{GPT-4o Mini} \\
\midrule
$U_{\mathrm{R+S+U}}$ & \textbf{0.41} & \textbf{0.35} \\
$U_{\mathrm{R+S}}$   & 0.39 & 0.35 \\
$U_{\mathrm{R+U}}$   & 0.35 & 0.33 \\
$U_{\mathrm{S+U}}$   & 0.35 & 0.31 \\
$U_{\mathrm{R}}$     & 0.32 & 0.33 \\
$U_{\mathrm{S}}$     & 0.29 & 0.29 \\
$U_{\mathrm{U}}$     & 0.30 & 0.28 \\
\bottomrule
\end{tabular*}
\vspace{-1em}
\end{table}

\begin{figure}[t]
    \centering
    \includegraphics[width=\linewidth]{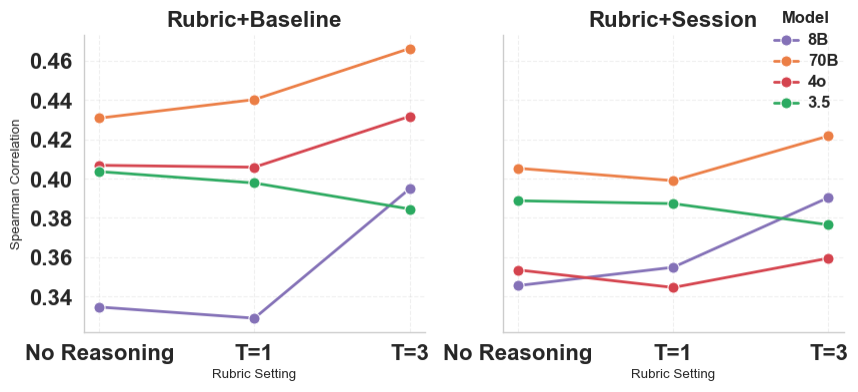}
    \caption{Effect of reasoning and rubric candidate size on usefulness prediction.}
    \label{fig:rubric_ablation}
\end{figure}

\subsection{Ablation of TRUE Rubric}
\label{sec:rubricablation}
We evaluate three rubric settings: No Reasoning, $T=1$, and $T=3$ (TRUE) on KDD'19 using TRUE$_{\text{Rubric+Baseline}}$ and TRUE$_{\text{Rubric+Session}}$ across LLaMA 3.3 70B, LLaMA 3.1 8B, GPT-3.5 Turbo, and GPT-4o Mini. No Reasoning uses direct labeling without manual rubric refinement, $T=1$ uses a single rubric candidate without manual refinement, and $T=3$ (TRUE) uses three reasoning-guided rubric candidates that are manually consolidated into a final rubric. For most models, increasing rubric candidate diversity from No Reasoning to $T=3$ improves usefulness prediction performance, with the largest gains observed for LLaMA 3.3 70B, LLaMA 3.1 8B, and GPT-4o Mini. However, GPT-3.5 exhibits the opposite trend, showing slightly lower correlations under both $T=1$ and $T=3$ than under No Reasoning. We hypothesize that GPT-3.5's weaker reasoning capabilities limit its ability to effectively utilize more complex reasoning-derived rubrics, resulting in performance degradation as rubric complexity increases. Nevertheless, the overall trend across the stronger models suggests that reasoning-guided rubric construction and rubric diversity contribute to improved usefulness assessment (shown in Figure~\ref{fig:rubric_ablation}).

\section{Discussion \& Limitations}
Our results show that LLMs can approximate usefulness judgments at or beyond the level of observed human agreement, supporting their viability as scalable proxies for user-centered evaluation. The gains from TRUE highlight the importance of structured reasoning: rubric-guided decomposition enables consistent integration of heterogeneous signals. We further find that relevance alone is insufficient to capture usefulness, particularly in session-based search where user goals evolve; incorporating task context and behavioral signals yields a more faithful representation of user utility. From a practical standpoint, TRUE enables scalable and reproducible usefulness labeling without requiring expensive human annotation or supervised training. Unlike LLM4Eval~\citep{rahmani_llm4eval_2024} and LLMJudge~\citep{rahmani_llmjudge_2024}, which focus on relevance and prompt-level aggregation, TRUE introduces a multi-stage rubric-driven framework that separates reasoning, rubric induction, and inference, improving interpretability and transferability. 

Despite these strengths, several limitations remain. First, rubric induction involves a structured manual consolidation step, which may introduce bias. Although TRUE induces its rubric from usefulness-labeled examples, we mitigate potential circularity by constructing the rubric on a held-out subset of KDD'19 and applying it unchanged to the remaining KDD'19 data as well as QRef and SIGIR16. The induced rubric remained stable across datasets, suggesting that it captures generalizable usefulness patterns rather than dataset-specific artifacts. Second, feature availability varies across datasets, requiring partial alignment of input signals. Third, while satisfaction signals can improve performance, they are not always available and are often better viewed as a downstream outcome of usefulness rather than a direct predictor. In our framework, satisfaction is treated as an auxiliary signal in an offline full-information setting rather than a required feature. Our comprehensive ablation results show that usefulness can still be reasonably inferred without satisfaction signals, although performance is generally strongest when multiple sources of contextual evidence are available. Fourth, user behavioral signals such as clicks, dwell time, and satisfaction are inherently noisy and incomplete proxies for usefulness, and may not fully capture user utility. Finally, LLM-based evaluation remains sensitive to prompt design and model choice, although TRUE mitigates this through structured rubrics, some variability persists. While this work focuses on improving agreement with human usefulness judgments, we do not directly evaluate downstream applications such as usefulness-aware ranking and system comparison. Future work will focus on reducing reliance on manual rubric design through fully label-agnostic approaches, on exploring fine-tuning strategies that better model user behavior signals, and on evaluating the impact of TRUE-generated usefulness labels in downstream IR applications.

\section{Conclusion}
Our work investigates the potential of LLMs for generating document usefulness labels in Web search evaluation. We introduce \textsc{TRUE}, a reproducible framework that models key dimensions of usefulness from a user’s perspective by incorporating session context, user behavior, and rubric-driven reasoning. Our results show that \textsc{TRUE} achieves moderate to strong correlation with human judgments, validating the effectiveness of the proposed approach. Unlike traditional machine learning methods, which often struggle to capture nuanced contextual relationships within search sessions, \textsc{TRUE} leverages the semantic reasoning capabilities of LLMs to provide a more user-aligned evaluation process. As a result, it consistently outperforms state-of-the-art ML baselines and existing LLM-based approaches across multiple datasets and settings. Given the inherent complexity of modeling usefulness, achieving strong alignment with human judgments represents a meaningful and practically valuable outcome. Rather than replacing relevance, we position usefulness as a complementary, user-centered evaluation signal that better reflects real search outcomes. This work contributes to the growing \textit{LLMs-as-Judges} paradigm~\citep{gu2024survey} by demonstrating how LLMs can effectively integrate search behavior signals for usefulness prediction. Future work will explore practical applications such as automated usefulness-based evaluation, offline and semi-online re-ranking, and search path recommendation. While our current work focuses on establishing reliable usefulness prediction, demonstrating strong alignment with human usefulness judgments is a necessary first step toward enabling these downstream applications. We therefore view usefulness-aware ranking, system comparison and retrieval optimization as important directions for future work.

\begin{acks}
This work was partially supported by cloud computing credits provided through an Amazon Research Award to Dr. Chirag Shah.
\end{acks}
% \newpage

\bibliographystyle{ACM-Reference-Format}
\bibliography{reference}

\end{document}